# GEXTEXT: DISEASE NETWORK EXTRACTION FROM BIOMEDICAL LITERATURE

Abstract

Rob O'Shea
robert.o'shea.17@ucl.ac.uk

# Contents

# Quick Facts

### What is already known
- Network models offer valuable insights into disease systems
- Extraction of these models from unstructured data is complicated by long-term dependency structures and implicit information.

### What this study adds
- We provide a method of probabilistic diseasome extraction from biomedical text corpora.
- We demonstrate that this method extracts latent information and estimates similarities beyond those mentioned in the text.

# Abstract

PURPOSE: We propose a fully unsupervised method to learn latent disease networks directly from unstructured biomedical text corpora. This method addresses current challenges in unsupervised knowledge extraction, such as the detection of long-range dependencies and requirements for large training corpora.

METHODS: Let $C$ be a corpus of $n$ text chunks. Let $V$ be a set of $p$ disease terms occurring in the corpus. Let $\boldsymbol{X} \in \{0,1\}^{n \times p}$ indicate the occurrence of $V$ in $C$. Gextext identifies disease similarities by positively correlated occurrence patterns. This information is combined to generate a graph on which geodesic distance describes dissimilarity. Diseasomes were learned by Gextext and GloVE on corpora of 100-1000 PubMed abstracts. Similarity matrix estimates were validated against biomedical semantic similarity metrics and gene profile similarity.

RESULTS: Geodesic distance on Gextext-inferred diseasomes correlated inversely with external measures of semantic similarity. Gene profile similarity also correlated significant with proximity on the inferred graph. Gextext outperformed GloVE in our experiments. The information contained on the Gextext graph exceeded the explicit information content within the text.

CONCLUSIONS: Gextext extracts latent relationships from unstructured text, enabling fully unsupervised modelling of diseasome graphs from PubMed abstracts.

# Introduction

Network models have become a popular paradigm to represent the complex domain of disease interaction and comorbidity (1–4). The flexibility of such models, termed "diseasomes" (2,3), facilitates the representation of complicated interactions between multiple biological phenomena. Diseasomes are typically created from bioinformatic databases, linking conditions according to shared characteristics, such as common therapies or gene associations (3,5,6). These models integrate established knowledge from structured databases of biological and clinical studies, curated with human expertise. Consequently, diseasome models typically assume a fixed structure according to the generating databases. The prospect of population-specific diseasome modelling is attractive, as it would allow the comparative analysis of disease systems under various states, potentially elucidating important functional characteristics. However, such projects would require structured bioinformatic databases made specifically for the target population, a resource which may be unattainable.

Recently, the prospect of unsupervised diseasome extraction from unstructured data has been proposed (7,8). A vast quantity of biomedical information is contained text data such as electronic health records (EHR) and scientific literature. Explicit and latent knowledge modelling in such corpora may provide new insights into disease pathogenesis and comorbidity interaction. Direct extraction of disease graphs from text sources would allow population-specific analyses and comparative insights. Mining of novel corpora, such as social media, may provide research access to hard-to-reach groups.

Unsupervised pattern-level knowledge modelling has followed two distinct approaches, distributed word representations and graphical relational models. Distributed word embedding models aim to map the corpus vocabulary to a set of real valued vectors such that vector-vector distances reflect the similarity of the corresponding term pair. Similarity is estimated probabilistically from term co-occurrence within short context windows. Word vector learning methods such as Continuous-Bag-Of-Words (9), Skip-Gram (9) and GloVE (10) enabled major developments in biomedical language understanding tasks (11–13). Open information extractions models (11,14–17) encode explicit relationships between entities as tuples. This exact approach is well suited to biomedical tasks, and result may easily be traced to supporting evidence in the text. Multiple relational statements may be compounded into a unified knowledge graph, supporting probabilistic prediction of unobserved relations (7,17).

Where references are implicit, long intervals of text may separate related entities. Such latent references present a challenge for knowledge extraction. Word embedding methods depend on term co-occurrence within a short context window. The context window may be extended, at the cost of including a greater proportion of uninteresting or unrelated terms, reducing the signal-to-noise ratio. Traditional word embedding methods generate a vector mapping for each term in the corpus vocabulary. In many knowledge extraction tasks, such as protein network inference, relationships are sought between a small subset of rare terms, which may co-occur infrequently in short context windows. The size of suitable corpora is often limited in such problems, complicating the embedding task. Open information extraction models depend on simple, correct and explicit descriptions of relationships. Consequently, incorrect statements, mutually conflictual statements and long-range patterns present challenges for this model class (14,18).

We provide a method to efficiently learn diseasome graphs from biomedical corpora in a probabilistic manner. This objective requires the discovery of implicit similarity information between disease terms, often across long context intervals and with small corpora. This similarity information is utilised to generate a disease network.

## Method

We present Gextext, a novel extraction technique, to address these challenges. Gextext ingests the corpus in large bag-of-words windows, facilitating the modelling of long-range dependencies. Disease relationships are identified by positively correlated term co-occurrence. This technique supports the embedding of a small set of disease terms, simplifying the knowledge representation objective. In this paper we present a technical description of the Gextext algorithm. We compare the performance of Gextext and GloVE in diseasome extraction from corpora of PubMed abstracts.

### Gextext Technical Overview

Let $V$ be a set of $p$ query terms. Let $G(V, E)$ denote the true knowledge graph of $V$, such that $E_{ij}$ indicates the existence of a relationship between the $i^{th}$ and $j^{th}$ terms in $V$. Let $C$ be a corpus of $n$ text chunks with vocabulary $W$. Let $\mathbf{X} \in \{0,1\}^{n \times p}$ be a binary matrix indicating the occurrence of $V$ in $C$, such that:

$$\mathbf{X}_{i,j} \leftarrow \begin{cases} 1, & V_j \in C_i \\ 0, & V_j \notin C_i \end{cases}$$

Let $\mathbf{\Sigma} \in \mathbb{R}^{\{p \times p\}}$ be the covariance matrix of $\mathbf{X}$, such that:

$$\mathbf{\Sigma}_{i,j} = \frac{\left(\mathbf{X}_{:,i} - \mu_{\mathbf{X}_{:,i}}\right)\left(\mathbf{X}_{:,j} - \mu_{\mathbf{X}_{:,j}}\right)}{n}$$

We define $\hat{G}$ according to the positive entries of $\mathbf{\Sigma}$, such that:

$$\hat{E}_{i,j} \leftarrow \begin{cases} 1, & \mathbf{\Sigma}_{i,j} > 0 \\ 0, & \mathbf{\Sigma}_{i,j} \leq 0 \\ 0, & i = j \end{cases}$$

Thus, Gextext infers semantic relationships between terms whose occurrence patterns are positively correlated.

### Task Overview

The tasks required the inference of a disease similarity matrix from a corpus of PubMed abstracts. Disease terms were extracted from the Human Disease Ontology (19). Various semantic similarity metrics have been described for pairs of biomedical terms, based on the graph structure of the ontology (20). These have been applied to the disease ontology via the software DOSim (21). We evaluate the quality of the inferred similarity matrices by correlation against these established metrics. We compare Gextext with GloVE, a static word embedding model (10). A biomedical text dataset was generated by downloading 10,000 PubMed abstracts with the search query "cancer AND gene". Abstracts returned by such a search were expected to contain piecemeal information on genes associated with cancer pathogenesis. Given the depth and breadth of this topic, it is expected that small corpora (100-1,000 abstracts) would explicitly state a small proportion of the available knowledge on the subject. Given $p$ disease terms occurring within the corpus, it is expected that explicit similarity information would be available for a small minority of disease pairs. Consequently, estimation of a complete similarity matrix for $\frac{p^2-p}{2}$ pairs would necessitate discovery of latent information within the corpus.

## Pre-Processing

Abstracts were downloaded in Medline format using the "easyPubMed" package. Task corpora were generated by randomly sampling either 100 or 1000 abstracts from this dataset. Each individual abstract was considered as a text chunk. Text chunks were tokenised three times to generate n-grams of size 1, 2 and 3. For each task $V$ was restricted to the disease ontology terms which occurred within the corpus.

## Gextext Similarity Inference

Each chunk was scanned for exact string matches with terms in $V$ to generate $X$. $\hat{G}$ was generated by the above method. A similarity matrix was generated according to the inverse of the geodesic distance between disease terms on $\hat{G}$, such that:

$$similarity_{\hat{G}}(V_i, V_j) \leftarrow \begin{cases} d_{\hat{G}}(V_i, V_j)^{-1}, & d_{\hat{G}}(V_i, V_j) < \infty \\ 0, & d_{\hat{G}}(V_i, V_j) = \infty \end{cases}$$

Where $d_{\hat{G}}(V_i, V_j)$ denotes the geodesic distance between $V_i$ and $V_j$ on $\hat{G}$.

## GloVE Similarity Inference

For comparison, term embeddings were learned on each task corpus with the GloVE algorithm. The method was implemented using the Text2Vec package at its default settings. 25, 50 and 100 dimensional embeddings were learned for each task. Context window was set to 5 terms. For each embedding, $Z$, the cosine similarity matrix of $Z$ was calculated between the elements of $V$, such that:

$$similarity_Z(V_i, V_j) \leftarrow \frac{Z_{V_i} \cdot Z_{V_j}}{\|Z_{V_i}\| \|Z_{V_j}\|}$$

## External Validation of Similarity Matrices

Similarity matrices inferred by Gextext and GloVE were validated against corresponding Wang (22), Resnik (23), Relevance (24), Jiang (25) and Lin (26) similarity measures. These measures define semantic similarity according to the structural features of the ontology graph. Descriptions of these measures are provided by Pesquita (20) and Jiang (21). Similarity matrices were calculated for $V$ using the DOSim package (21). To evaluate the concordance of the graph with the true biological similarity, we calculated a further measure of disease similarity based on gene profile overlap. Gene associations were extracted from the disease ontology for each term in $V$ (27). The Jaccard similarity of these profiles was measured such that:

$$similarity_{GeneOverlap}(V_i, V_j) \leftarrow \frac{GeneProfile_i \cap GeneProfile_j}{GeneProfile_i \cup GeneProfile_j}$$

# Results

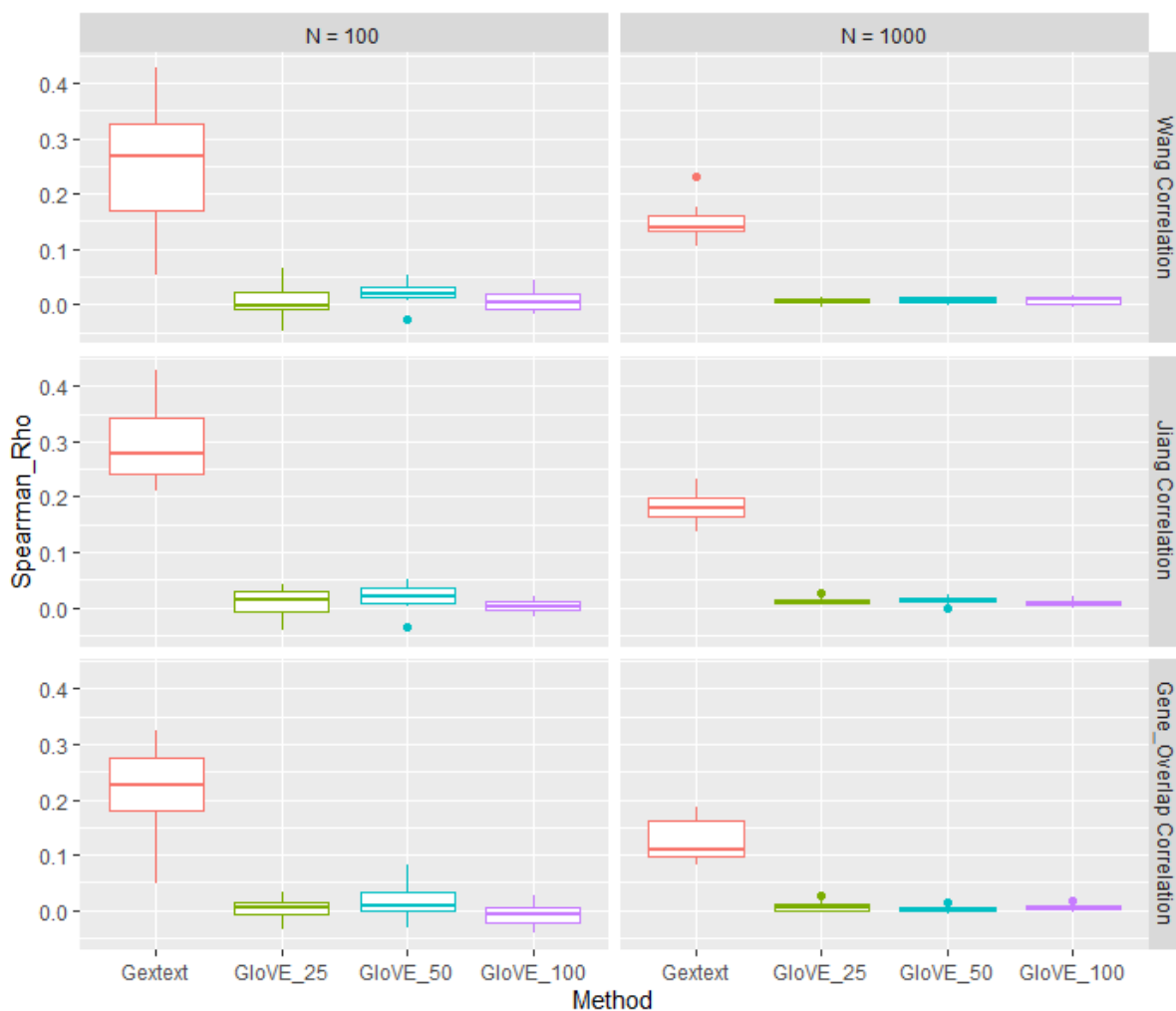

*Figure 1: Performance of Gextext and GloVE in the 100 abstracts tasks (left) and 1000 abstracts tasks (right). Each task required the estimation of a similarity matrix for a vocabulary of disease terms, given a corpus of PubMed abstracts. Similarity matrices generated by each method were compared with Jiang and Wang semantic similarity metrics by Spearman correlation. Similarity matrices were also compared to the Jaccard similarity of gene profiles associated with each disease term (Gene_Overlap).*

## Gextext Performance

In the 100 abstracts task, approximately 60 disease terms were identified per corpus (mean = 60, range = [49, 77]). Disease graphs learned by Gextext were sparse (mean = 0.062, range = [0.049, 0.076]). A small proportion of vertices were isolated in these graphs (mean = 0.086, range = [0.056, 0.169]). Gextext graphs reflected external similarity metrics. In the 100 abstracts task, Gextext similarity correlated significantly with the Wang (Spearman Rho = 0.254, p <2e-16), Resnik (Spearman Rho = 0.238, p <2e-16), Relevance (Spearman Rho = 0.283, p <2e-16), Jiang (Spearman Rho = 0.293, p <2e-16) and Lin (Spearman Rho = 0.285, p <2e-16) similarity measures. Furthermore, Gextext similarity correlated significantly with the pathogenetic overlap between diseases (Spearman Rho = 0.219, p <2e-16).

In the 1000 abstracts task, approximately 242 disease terms were identified (mean = 239.8, range = [224, 267]) per corpus. These graphs demonstrated higher sparsity (mean = 0.02, range = [0.019, 0.022]) than those of the 100 abstracts task. A smaller proportion of isolated vertices was observed in this task than the 100 abstracts task (mean = 0.082, range = [0.064, 0.121]). Again, Gextext similarity correlated significantly with the Wang (Spearman Rho = 0.148, $p < 2e-16$), Resnik (Spearman Rho = 0.191, $p < 2e-16$), Relevance (Spearman Rho = 0.2, $p < 2e-16$), Jiang (Spearman Rho = 0.181, $p < 2e-16$) and Lin (Spearman Rho = 0.2, $p < 2e-16$) measures. Gextext similarity correlated significantly with the pathogenetic genetic overlap between diseases (Spearman Rho = 0.128, $p < 2e-16$).

### GloVE Performance

The embeddings learned by GloVE were less informative than those of Gextext. In the 100 abstracts task, the 50 dimensional embedding delivered the best GloVE performance, though cosine similarities correlated poorly with the Wang (Spearman Rho = 0.02, p 0.19), Resnik (Spearman Rho = 0.016, p 0.48), Relevance (Spearman Rho = 0.019, p 0.17), Jiang (Spearman Rho = 0.019, p 0.13), Lin (Spearman Rho = 0.019, p 0.16) or Gene Overlap (Spearman Rho = 0.016, p 0.015) similarity measures. GloVE also failed to generate informative embeddings in the 1000 abstracts task. Again, the 50 dimensional embeddings performed best, yet correlated weakly with the Wang (Spearman Rho = 0.007, p 0.0064), Resnik (Spearman Rho = 0.008, p 0.00027), Relevance (Spearman Rho = 0.009, p 0.00011), Jiang (Spearman Rho = 0.012, p 4.6e-07), Lin (Spearman Rho = 0.009, p 1e-04) and Gene Overlap (Spearman Rho = 0.004, p 0.14) similarity measures.

| Method | N Abstracts | Wang | Resnik | Relevance | Jiang | Lin | Gene Overlap |
|---|---|---|---|---|---|---|---|
| Gextext | N = 100 | 0.254 | 0.238 | 0.283 | 0.293 | 0.285 | 0.219 |
| GloVE_25 | N = 100 | 0.004 | 0.003 | 0.005 | 0.007 | 0.004 | 0.003 |
| GloVE_50 | N = 100 | 0.02 | 0.016 | 0.019 | 0.019 | 0.019 | 0.016 |
| GloVE_100 | N = 100 | 0.006 | 0.011 | 0.014 | 0.001 | 0.014 | -0.008 |
| Gextext | N = 1000 | 0.148 | 0.191 | 0.2 | 0.181 | 0.2 | 0.128 |
| GloVE_25 | N = 1000 | 0.006 | 0.008 | 0.009 | 0.013 | 0.009 | 0.008 |
| GloVE_50 | N = 1000 | 0.007 | 0.008 | 0.009 | 0.012 | 0.009 | 0.004 |
| GloVE_100 | N = 1000 | 0.007 | 0.009 | 0.009 | 0.008 | 0.009 | 0.006 |

*Table 1: Mean Spearman correlation between inferred similarity matrices and true similarity measures. Wang, Resnik, Relevance, Jiang and Lin measures define similarity in terms of the Disease Ontology graph. "Gene_Overlap" similarity was defined as the Jaccard similarity of the gene profile of each disease.*

### Gextext Latent Information Performance

Gextext gathered information on disease similarity in a pairwise fashion, but combined these relationships to estimate the entire similarity matrix of V. In the 100 abstract tasks, a small minority of disease term pairs co-occurred within the dataset (mean = 0.062, range = [0.048, 0.091]). Similarity estimates for the remaining 92% of term pairs were estimated indirectly. Approximately 1% of disease terms co-occurred in the 1,000 abstracts tasks (mean = 0.025, range = [0.02, 0.029]) necessitating the reconstruction of similarity information for an overwhelming majority of disease pairs. A second task evaluated the disease pairs which did not co-occur in the corpus. In the 100 abstracts task, geodesic distance between non-co-occurring pairs correlated with the Wang (Spearman Rho = 0.228, $p < 2e-16$), Resnik (Spearman Rho = 0.232, $p < 2e-16$), Relevance (Spearman Rho = 0.245, $p < 2e-16$), Jiang (Spearman Rho = 0.184, $p < 2e-16$), Lin (Spearman Rho = 0.244, $p < 2e-16$) and Gene_Overlap (Spearman Rho = 0.181, $p < 2e-16$) measures. Latent similarities learned in the 1000 abstracts tasks also correlated with Wang (Spearman Rho = 0.104, $p < 2e-16$), Resnik (Spearman Rho = 0.13, $p < 2e-16$), Relevance (Spearman Rho = 0.135, $p < 2e-16$), Jiang (Spearman Rho = 0.122, $p < 2e-16$), Lin (Spearman Rho = 0.134, $p < 2e-16$) and Gene_Overlap (Spearman Rho = 0.096, $p < 2e-16$) measures.

## Discussion

### Performance Comparison

In our analysis, graphs generated by the Gextext method outperformed GloVE embeddings by all measures. Geodesic distance on the Gextext inferred graph correlated significantly with semantic similarity of disease terms. This indicates that Gextext learned a coherent map of related disease terms from the corpora in each task. Furthermore, the concordance of geodesic distance with the similarity of disease gene profiles indicates that Gextext also provided a useful biological model of disease similarity. These results confirm that Gextext discovered latent information on disease similarities within the text.

It should be noted that GloVE returns an embedding which contains a vector representation for each term in the corpus vocabulary, scaling as $O(|W|^2)$. In comparison, Gextext returns an embedding for a small target vocabulary $V$, scaling as $O(|V|^2)$. In our applications $|W|$ greatly exceeded $|V|$, simplifying the inference problem. This simplification sacrifices some of the richness of the model, under the assumption that the excluded terms are uninformative with respect to disease similarity. Seemingly arbitrary terms may provide information on disease similarity, for example, the name of an endemic region. However, in our experiments it was observed that prioritisation of a target vocabulary resulted in a superior model of these terms. Furthermore, subsetting the target vocabulary removes the vast majority of noise terms, allowing expansion of the context window to abstract length. This allowed extraction of long-range dependencies from the corpus.

### Synergistic Information Modelling

The majority of similarity information estimated by in these tasks compared diseases which did not co-occur in the corpus. This required the inference of indirect relationships and implicit information. Gextext's successful estimation of such similarities demonstrates that the method is models latent information in the text. This allowed the characterisation of large disease networks from small corpora of randomly selected PubMed abstracts.

### Experimental Observations

The correlation between geodesic distance on the Gextext graph and semantic similarity was greater in the 100 abstracts task than the 1000 abstracts task. This phenomenon is probably a consequence of the increased proportion of rare disease terms in $V$ as the corpus size increases. Given the small number of occurrences of such topics within the corpus, little information is available to model their similarity profile within the corpus. Given that all terms are more likely to occur with common terms than uncommon terms, the graph is expected to grow in an approximately scale-free distribution pattern (28,29). A consequence of this phenomenon is that paths between rare disease terms are typically indirect, passing through more common nodes. Accordingly, the majority of explicit similarity information should describe relations to common terms. Thus, similarities between infrequent terms are modelled with almost exclusively implicit information, resulting in a less reliable estimate.

In applications of Gextext to model disease interaction graphs for a rare disease, corpora should be collected to specifically include the disease with high frequency. It is expected that Gextext will reliably model the relationships between the rare disease and many common diseases. However, relationships modelled between infrequent terms should considered carefully, and examined with validation corpora.

## Conclusions

Gextext efficiently extracts latent relationships from unstructured text, enabling reliable characterisation of disease similarity graphs. This method supports the estimation of disease similarity matrices based on a small corpus of unstructured biomedical text. These knowledge models contain information on semantic and biological disease similarity based on the information in the given corpus. Provision of a corpus of documents relating to an arbitrary rare disease will facilitate the estimation of its diseasome. This is a rapid approach to generate a rich comorbidity profile estimate for a rare condition. Furthermore, classical statistical network analysis of such disease models may identify systematic vulnerabilities of the graph, indicating optimal preventative and therapeutic strategies.

An alternative application of Gextext is the comparative analysis of diseasomes in multiple corpora. For example, diseasomes may be learned on electronic health records of two populations. In such corpora, diseasomes would represent comorbidity risk networks. Comparative evaluation of such networks may identify complex causative factors for health discrepancies between the two groups.